\documentclass[aps,prb,amsmath,amssymb]{revtex4}

\usepackage{graphicx}
\usepackage{amssymb}

\newcommand{\deff}{{\tt def}}
\newcommand{\inc}{{\tt inc}}
\newcommand{\divv}{{\tt div}}
\newcommand{\curl}{{\tt curl}}
\newcommand{\grad}{{\tt grad}}

\begin{document}

\title{Einstein-Cartan theory as a theory of defects in space-time}

\author{M. L. Ruggiero}
\email{matteo.ruggiero@polito.it}
\author{A. Tartaglia}
\email{angelo.tartaglia@polito.it} \affiliation{Dip.\ Fisica,
Politecnico and INFN, Torino, Italy, I-10129 }

\begin{abstract}
The Einstein-Cartan theory of gravitation and the classical theory
of defects in an elastic medium are presented and compared. The
former is an extension of general relativity and refers to
four-dimensional space-time, while we introduce the latter as a
description of the equilibrium state of a three-dimensional
continuum. Despite these important differences, an analogy is
built on their common geometrical foundations, and it is shown
that a space-time with curvature and torsion can be considered as
a state of a four-dimensional continuum containing defects. This
formal analogy is useful for illustrating the geometrical concept
of torsion by applying it to concrete physical problems. Moreover,
the presentation of these theories using a common geometrical
basis allows a deeper understanding of their foundations.
\end{abstract}

\maketitle

\section{Introduction}
The Einstein-Cartan Theory (ECT) is a gravitational theory whose
observational predictions are in perfect agreement with the
classical tests of general relativity. From a geometrical point of
view, ECT turns out to be an extension of general relativity to
Riemann-Cartan spaces in which both the metric and the
\textit{torsion} determine the geometry of space-time.

As we shall see, torsion refers to the non-symmetric part of the
affine connection in a manifold, and in general relativity the
torsion is assumed to be zero. However, the introduction of
torsion is not merely a mathematical trick to obtain a richer
geometrical structure. From a physical point of view, torsion in
ECT is generated by the spin. Hence, in ECT, both mass and spin,
which are intrinsic and fundamental properties of matter,
influence the structure of space-time.

In non-gravitational physics the concept of torsion is well known.
For instance, in simple supergravity non-symmetric connections are
used to handle particles and fields.\cite{west90} However, we want
to introduce the concept of torsion by illustrating its use in the
theory of defects, where curvature and torsion are used to
describe the geometric properties of a material continuum.

We shall point out that the Einstein-Cartan theory of gravitation
and the theory of defects have similar fundamental equations, and
we shall stress the analogies and differences in their underlying
geometric structure. Because the two theories describe very
different physical phenomena, it is clear that the comparison can
only be formal. However, we believe that this way of presenting
the subject will give deeper insight into the geometrical tools
used in both theories.

We present the fundamental equations of the theory of defects in
Sec.~\ref{sec:TD} and give their geometrical interpretation in
Sec.~\ref{sec:geometry}. In Sec.~\ref{sec:ECT} we introduce ECT as
a viable theory of gravitation. Then, in Sec.~\ref{sec:TDECT} we
compare the two theories and also briefly illustrate some possible
applications of the theory of defects to contemporary physics such
as the study of cosmic strings.

\section{Defects in a continuous medium}\label{sec:TD}
The theory of defects\cite{footnote1} deals with bodies that are
subjected to internal stresses even if there are no external
forces acting on them. These stresses are caused by the presence
of defects, which are alterations of the ideal and ordered
structure of the medium that rearrange its structure to reach
internal equilibrium. In particular, we shall always refer to the
static version of the theory of defects.

We shall adopt a continuum description of a medium with defects,
which allows a geometric approach that is suitable for a
comparison with ECT. We shall refer to a continuum ensemble of
Bravais lattices, which is obtained by a limiting process that
transforms the crystal into a continuum. The (ideal) limiting
process is performed in such a way that the ``mass points'' are
split into smaller particles, which are arranged so that in any
given volume $\Delta V$, the same type of lattice, only with
smaller spacing, is preserved. Moreover, we require that both the
mass content and the defect content in $\Delta V$ remain
unchanged, so that the mass density and defect density can be
defined. Note that in this process the continuum preserves the
crystallographic directions.

This continuum approach neglects the details of the interactions
that characterize the body on the microscopic scale, but is
suitable for describing large scale phenomena. In a continuum
theory, the properties of the body are given in terms of just a
few physical quantities. Mathematically these quantities are
functions that may have discontinuities, but only a few so as to
maintain the macroscopic continuity of the body even when it
contains defects.  But, what is a defect?

\begin{figure}[h]
\begin{center}
\includegraphics[width=4.5cm,height=4.5cm]{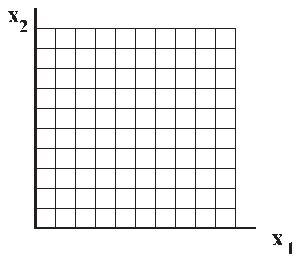}
\caption{Two-dimensional representation of the undeformed volume
element $\Delta V$ and its lattice structure.} \label{fig:kro230}
\end{center}
\end{figure}

It is useful to introduce the idea of deformation.\cite{bib:kron}
Consider a small volume element $\Delta V$ (see
Fig.~\ref{fig:kro230}, where the volume element $\Delta V$ is
sketched with its lattice structure only to give a more intuitive
description). If we assign to its points cartesian coordinates
$(x^1,x^2,x^3)$, we can make a correspondence between the
coordinates of a physical point $P$ before the deformation and its
coordinates after the deformation denoted by a
prime\cite{footnote2}:
\begin{equation}
x{'}^i=x^i+u^i(x) . \label{eq:def1}
\end{equation}
The deformation is represented by the displacement field $u^i(x)$.
It is easy to show that if $dl$ is the distance between the two
points before the deformation, then after the deformation, $dl'$
is given by\cite{bib:landa7}
\begin{equation}
{dl'}^2=dl^2+2\varepsilon_{ij}dx^i dx^j, \label{eq:def2}
\end{equation}
where $\varepsilon_{ij}$ is the strain tensor
\begin{equation}
\varepsilon_{ij}(x) = \frac{1}{2}(\partial_i u_j + \partial_j u_i
+
\partial_i u_l \partial_j u^l). \label{eq:strain0}
\end{equation}
We shall consider bodies for which the deformations are ``small,''
and we shall use for the strain tensor the expression
\begin{equation}
\varepsilon_{ij}(x) = \frac{1}{2}(\partial_i u_j +\partial_j u_i
). \label{eq:strain}
\end{equation}
Thus, we confine ourselves to a linear theory in which both the
displacement field and its derivatives are small enough to neglect
quadratic terms.

Because we shall use a different approach in Sec.~\ref{sec:TD}, we
want to stress that here the positions of the physical points of
the bodies, both before and after the deformation, are referred to
a fixed cartesian system of coordinates.

\begin{figure}[h]
\begin{center}
\includegraphics[width=10cm,height=4.5cm]{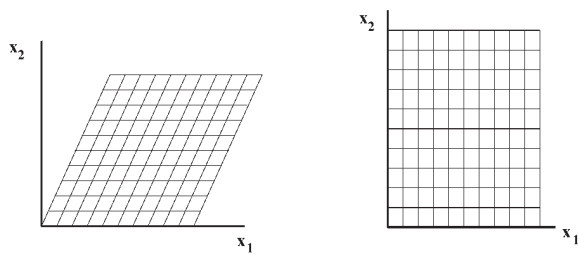}
\caption{two-dimensional representations of elastic deformations,
which result from tilt and stretching operations on the
underformed volume element. Parts of the body that were close to
each other remain close even after the deformation.}
\label{fig:kro231}
\end{center}
\end{figure}

Let us introduce the concept of plastic and elastic deformation.
In an elastic deformation parts of the body that are close to each
other remain close even after the deformation. Such a deformation
is shown in Fig.~\ref{fig:kro231}, where the lattice which
represents the ``particles'' of the body is dragged along with the
volume.

\begin{figure}[h]
\begin{center}
\includegraphics[width=10cm,height=8cm]{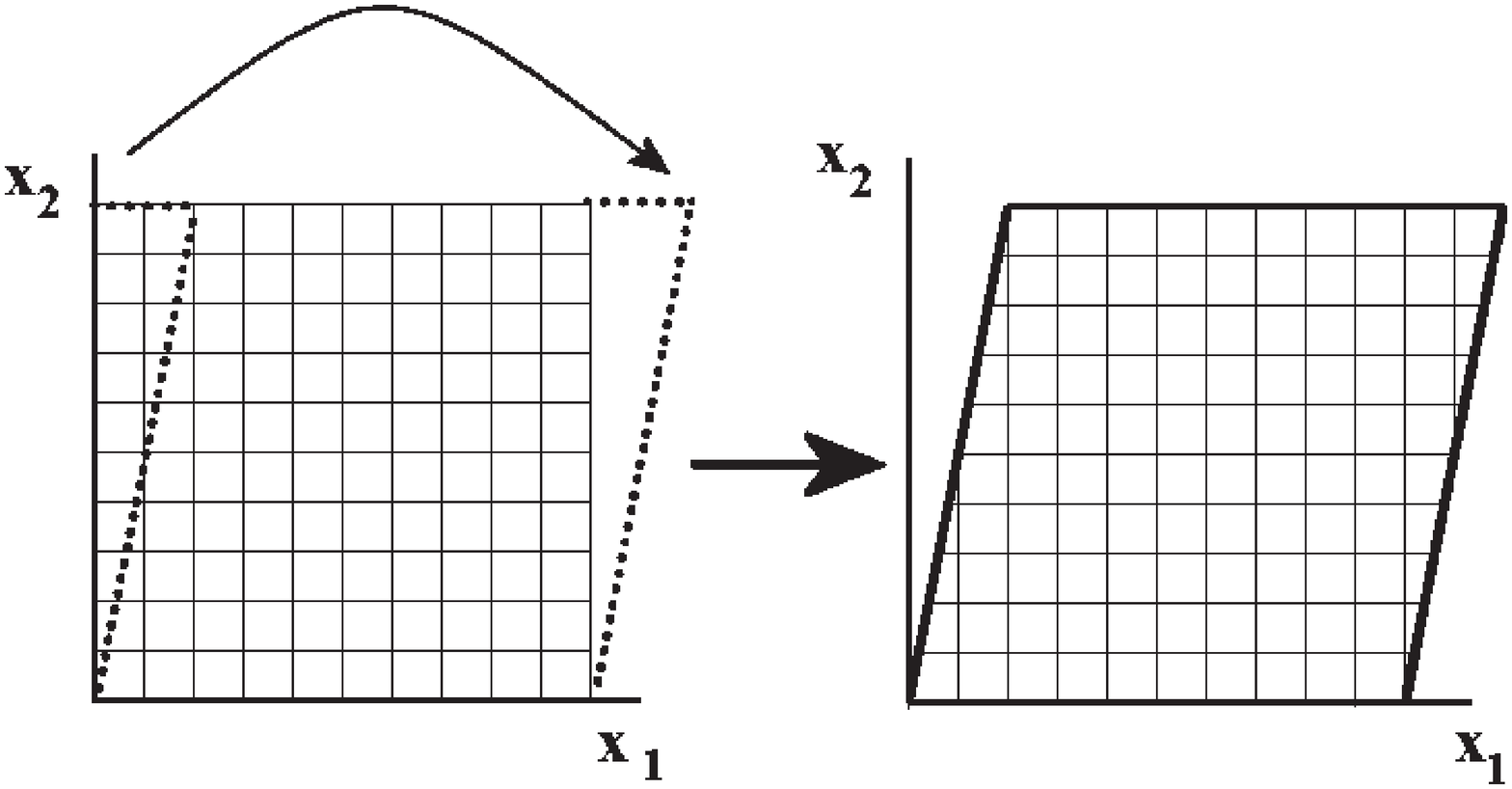}
\caption{Two-dimensional representation of a plastic deformation,
in which a piece of the undeformed volume is cut and then pasted
in a different position.} \label{fig:kro232bis}
\end{center}
\end{figure}

A plastic deformation can be described by the ``cut and paste''
procedure depicted in Fig.~\ref{fig:kro232bis}, where pieces of
matter are removed from one side and pasted on the opposite side
of $\Delta V$ to obtain a compact but deformed volume element. We
see that this procedure does not change the lattice structure. In
this process pieces have been moved from one place to another in
the volume, so that they have new neighboring pieces.

The use of this procedure was introduced by
Volterra\cite{bib:volterra} (quoted by Nabarro in
Ref.~\onlinecite{nabarro67}) in a paper devoted to the study of
elastic deformations of multiply connected, three-dimensional
bodies. The different constructions of the Volterra procedure lead
to different kinds of defects. These ideal procedures can be
performed on all the elementary volumes $\Delta V$ that compose
the body. After the deformation, we may have two different
situations: the deformed elements fit perfectly or they do not. In
the latter case, we must apply a further deformation to re-compact
the body. In the first case, we speak of a compatible deformation,
in the second we have an incompatible deformation. Naively, a body
remains compact (non-compact) if the deformation is compatible,
while it changes its compactness (connectedness) state, if it
undergoes an incompatible deformation.

Mathematically, an incompatible deformation corresponds to the
non-integrability of the differential form $du^i$, which means
that the displacement field $u^i(x)$ is multivalued, and thus that
discontinuities arise when passing from one elementary volume
element to another. This fact is expressed by,
\begin{equation}
\epsilon^{ikl}\partial_k \partial_l u^j = \alpha^{ij} \neq 0,
\label{eq:incomp1}
\end{equation}
where $\epsilon^{ijk}$ is the completely antisymmetric tensor of
rank three.

On the other hand, when the field is integrable, we have
$\epsilon^{ikl}\partial_k \partial_l u^j =0$, and the deformation
is described by a displacement field well defined in the entire
body. Equation~(\ref{eq:incomp1}) says that a body is in an
incompatible state when there is a ``source of incompatibility''
as represented by the tensor $\alpha_{ij}$ (see
Ref.~\onlinecite{bib:kron}).

These arguments help us understand the concept of a defect. The
Volterra processes are just mental experiments, useful to explain
and visualize incompatibility. A body may exist in an
incompatibility state provided that there are defects in its
structure, which are the sources of the incompatibility. For
instance, the $\alpha_{ij}$ tensor appearing in
Eq.~(\ref{eq:incomp1}) is called the density of dislocations,
which are defects as described below.

We want to stress that the incompatibility state of the body
should not be thought of as a result of a deformation used as a
visualization tool. Rather, the incompatibility state is the
result of the presence of defects in the structure of the body. In
this section and the following ones, we shall often speak of
incompatible deformations that correspond to a ``defect state'' of
the medium. Nevertheless, this defect state should not be
interpreted to be the result of an action on the body by an
external agent. The concept of an incompatible deformation has
been introduced because it is useful to understand the effects of
the presence of defects. That is, when a body is filled with
defects, its structure is similar to the one that can be reached
due to the cut and paste procedure which we have introduced.

\begin{figure}[h]
\begin{center}
\includegraphics[width=12cm,height=12cm]{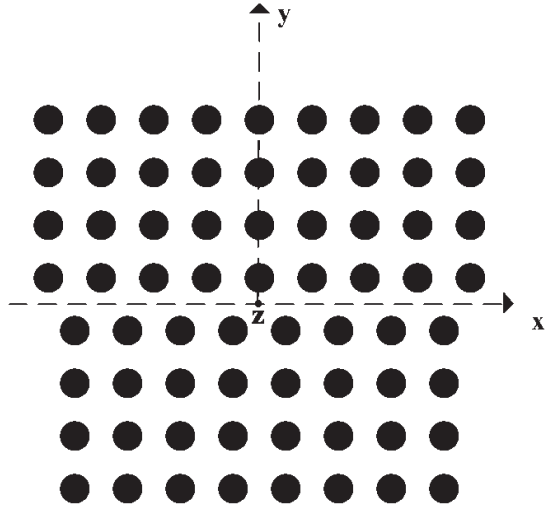}
\caption{An ``extra'' half plane is inserted in the lattice, whose
cross section is represented; the $z$ axis coincides with the line
of the edge dislocation.} \label{fig:dislocazione}
\end{center}
\end{figure}

\begin{figure}[h]
\begin{center}
\includegraphics[width=10cm,height=6cm]{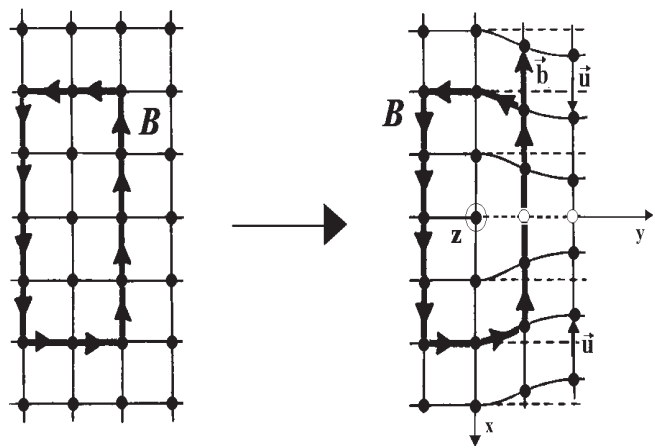}
\caption{Burgers vector: on the left a path $B$ in the undeformed
medium and, on the right, the correspondent path $B$ in the
deformed medium: for each step along a lattice direction in the
undeformed medium, a corresponding step is made in the deformed
medium. The $z$ axis is the dislocation line.} \label{fig:Burgers}
\end{center}
\end{figure}

We now will introduce in more detail some examples of defects and
their corresponding mathematical characterization. The essential
idea of a dislocation is geometric, and it can be illustrated by a
simple argument.\cite{bib:landa7} Let us imagine that an extra
half plane is inserted into a lattice whose cross section is
represented in Fig.~\ref{fig:dislocazione}. The $z$ axis coincides
with the edge of the half plane that has been introduced and is
called an edge dislocation (the $z$ axis is called a dislocation
line). In the neighborhood of the dislocation, the lattice planes
fit together in an almost regular manner. The presence of defects
causes a rearrangement of the whole structure, and the
deformations exist even far from the dislocation. This far
reaching rearrangement  is easily evidenced if we consider any
closed circuit of lattice points that encircles the dislocation
line. If $\mathbf{u}$ is the displacement field that denotes the
displacement of each point from its position in the ideal lattice,
the total increment of this vector field around a circuit will not
be zero, but it will be equal to a lattice spacing. Even though we
have considered an edge dislocation only, the change in the
displacement field is a general property of dislocations. After a
round trip along any closed circuit that encloses a dislocation
line, the displacement field increases by a vector which is called
the Burgers vector of the dislocation. We can characterize
mathematically the Burgers vector using Fig.~\ref{fig:Burgers},
where an edge dislocation, which corresponds to removing a layer
of lattice points, is depicted.\cite{footnote3} On the left, we
have a closed path in the undeformed medium; on the right, the
same path is traced in the deformed structure and shows the
Burgers vector due to the presence of a dislocation. It is evident
that, although the former path is closed, the latter is not. The
deformation is defined by Eq.~(\ref{eq:def1}), which we write in
discrete form
\begin{equation}
\mathbf{x}_{n}^{\prime}=\mathbf{x}_{n}+\mathbf{u}(\mathbf{x}_{n}),
\end{equation}
where $n$ labels the nodes of the lattice. The difference between
the position of the material points, as a consequence of a small
deformation, can be written in the form,
\begin{equation}
\delta \mathbf{x}_n= \delta \mathbf{u}_n, \label{eq:deltau}
\end{equation}
as a function of the small displacement field $\delta
\mathbf{u}_n$.

The closure gap is characterized by the Burgers vector $b^i$,
which is defined by,
\begin{equation}
\sum_{n=1}^N\delta u^{i}_n=\sum_{B}\delta u^{i}_n=b^{i},
\label{eq:Burdis}
\end{equation}
where $N$ is the total number of steps in the closed undeformed
loop $B$ and $\delta u^i_n$ refers to the value of the
displacement field in each step (see
Ref.~\onlinecite{bib:klein89}).

The mathematical meaning of the Burgers vector becomes clear when
we pass to the continuous limit by letting the lattice spacing go
to zero.  In this limit Eq.~(\ref{eq:Burdis}) becomes
\begin{equation}
\oint_{B}du^{i} \equiv \oint_B \frac{\partial u^i}{\partial x^j}
dx^j =b^{i}. \label{eq:Burcon}
\end{equation}
Equation~(\ref{eq:Burcon}) makes an explicit correspondence
between the integrability properties of the differential form
$du^i$ when dislocations are present and the Burgers vector and is
independent of the structure of the lattice within the path $B$.
The Burgers vector depends only on the fact that the path encloses
a defect line.\cite{bib:kron}

It is easy to see the dependence of the Burgers vector on the
density of dislocations. If we use Eqs.~(\ref{eq:incomp1}) and
(\ref{eq:Burcon}) and Stokes' theorem, we obtain
\begin{equation}
db^h=\alpha^{nh} dA_n, \label{eq:dbalpha2}
\end{equation}
where $dA^n$ is the axial vector of the infinitesimal area element
enclosed by the path $B$.

The structure of the medium is locally undeformed and to detect
the presence of defects, a Burgers circuit is needed, which is a
global procedure because it requires measurements in different
places in the medium. The same holds for disclinations (see
below). This fact will have an interesting interpretation when we
describe defects in terms of geometrical properties, introducing
torsion and curvature. By analogy, these are not seen locally, but
appear when parallel transporting vectors in a differential
manifold endowed with a suitable affine connection. Indeed,
locally, the geometry always can be chosen to be flat.

\begin{figure}[h]
\begin{center}
\includegraphics{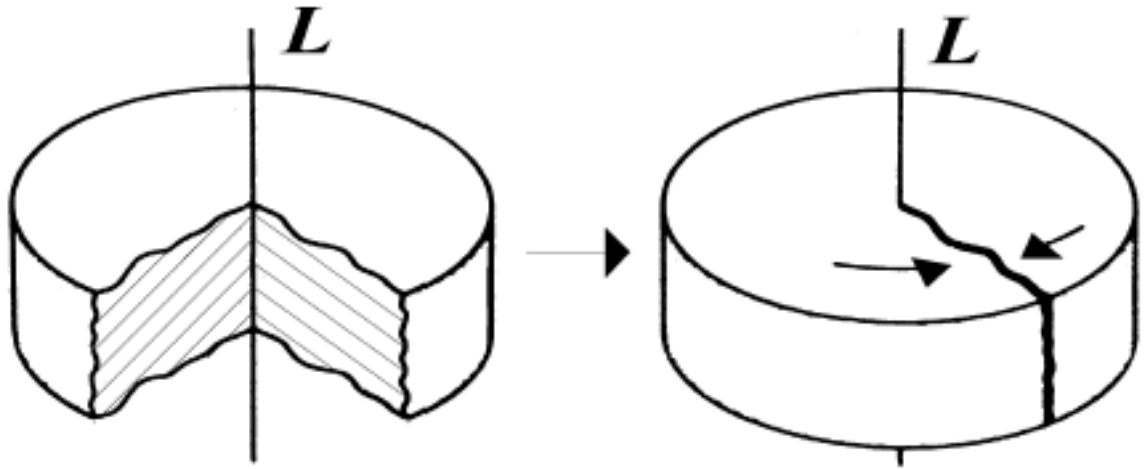}
\caption{We obtain a wedge disclination removing a slice from the
disk-like piece of medium, and then uniting the opposite faces
along the cut. $L$ is the disclination line.} \label{fig:kle787}
\end{center}
\end{figure}

Dislocations break the translational symmetry of the medium, while
disclinations break its rotational symmetry. It is easy to
understand what a disclination is if we perform the following
Volterra process. Start from the disk-like medium depicted in
Fig.~\ref{fig:kle787}. We remove a small slice, and then we unite
the opposite faces along the cut. In this way, we obtain a medium
that has an ordered structure everywhere except along the cut
surface (the medium appears locally undeformed), where the
displacement field is non-continuous. Mathematically a
discontinuity appears, which corresponds to a wedge disclination.

The discontinuity $\Delta u^i$ of the displacement field is
characterized by introducing a rotation vector $\Omega^{i}$
(Frank's vector) parallel to the line $L$:
\begin{equation}
\Delta u^{i}(\mathbf{x})=(\mathbf{\Omega}\wedge \mathbf{x})^{i}.
\label{eq:rotinf}
\end{equation}
This description is meaningful if the wedge is small, that is, if
we adopt a linear theory. For disclinations, the "linearity"
condition is a very important issue, because if we want to
introduce a density of disclinations and the corresponding Frank
vectors, we must be able to sum these vectors to obtain the
resultant one. As is well know, we cannot sum finite rotations in
a straightforward way, because they do not commute. So we can
define a density of disclinations provided that we confine
ourselves to a linear theory, where the rotations described by the
Frank's vector are infinitesimal and can be summed without
ambiguity.

The disclination density corresponding to each line is given by
\begin{equation}
\Theta_{ij}(\mathbf{x})=\delta_{i}(L)\Omega_{j},
\label{eq:discfrank}
\end{equation}
where $\delta_i(L)$ is a Dirac delta function which is nonzero
only along the disclination line $L$.\cite{bib:klein89}

Volterra processes lead to six kinds of configurations, which
belong to the six degrees of freedom of the proper group of motion
in $\mathbb{R}^3$, that is, the Euclidean group defined by the
semidirect product $SO(3)\otimes_S T(3)$. Hence, deformations
resulting from the translational subgroup $T(3)$ are dislocations,
while those belonging to the rotational subgroup $SO(3)$ are
disclinations. As we shall see, this construction can be extended
from 3 to $3+1$ dimensions, that is, by passing from Euclidean
space to space-time.

Now we can link the concept of incompatibility to the density of
defects in the medium. The presence of defects modifies the
geometry of the medium. For example, as we see in
Eq.~(\ref{eq:def2}), the distance between two physical points is
different when a deformation is present. Moreover, we know that a
deformation is the manifestation of the defects inside the medium,
which corresponds to an incompatibility state. De St.\
Venant\cite{bib:dsv} (quoted in Ref.~\onlinecite{bib:kron}; see
also Ref.~\onlinecite{tt} for a thorough review) showed that when
a medium is in a state of compatible deformation, the following
condition is fulfilled by the strain tensor $\varepsilon_{ij}$
\begin{equation}
\epsilon^{ikm}\epsilon^{jln}\partial_k \partial_l
\varepsilon_{mn}=0. \label{eq:dSV}
\end{equation}
In this case, the differential form $du^i$ is integrable, the
displacement field $u^i(x)$ is well defined in the entire medium,
and the strain tensor is its symmetric gradient (see
Eq.~(\ref{eq:strain})). On the other hand, when defects are
present, Eq.~(\ref{eq:dSV}) becomes
\begin{equation}
\epsilon^{ikm}\epsilon^{jln}\partial_k \partial_l
\varepsilon_{mn}=\chi^{ij}, \label{eq:incII}
\end{equation}
and the strain is said to be incompatible with the defects. The
latter are characterized by the total density of defects tensor
$\chi^{ij}$; moreover, $du^i$ is not integrable, and the
displacement field is multivalued. The density of defects
$\chi^{ij}$ is expressed in terms of the density of dislocations
$\alpha_{ij}$ and the disclinations $\Theta_{ij}$ by the following
relation (see, for example, Ref.~\onlinecite{bib:klein89}):
\begin{equation}
\chi^{ij} \equiv \Theta^{ji} - \epsilon^{jmn}\partial_m (
-\alpha^{i}_{n}+\frac{1}{2}\delta^{i}_{n}\alpha^k_{k}).
\label{eq:chi}
\end{equation}

By means of eqs.~(\ref{eq:dSV}), (\ref{eq:incII}), (\ref{eq:chi})
we have outlined the mathematical definition of defects. However,
Eqs.~(\ref{eq:dSV})--(\ref{eq:incII}) also have a more intuitive
and physical interpretation. The total density of defects,
$\chi_{ij}$, influences the state and the geometric properties of
the medium. So we can think of $\chi_{ij}$ as a source term in a
constitutive equation. In the following we shall see how this
approach and the related definitions  correspond to ECT.

For our purposes, we are interested in the geometrical description
of the equilibrium state of a medium filled with defects, and we
shall not study how stresses are linked to strains by a
generalized Hooke's law. In other words, we shall not consider the
elastic reaction of the medium to the presence of defects.
However, this problem is very interesting and consists in
determining the stresses and strains of a given medium when the
distribution of defects in it is given. It is worth mentioning
that this internal problem of equilibrium is solved by a very
elegant theory\cite{bib:kron58} that is based on the analogy
between electromagnetism and the theory of defects. In Appendix~B
we shall describe this analogy in more detail, because we think
that it would be of pedagogical interest to compare the
mathematical structure of the fundamental equations of the two
theories.

\section{Geometry of Defects}\label{sec:geometry}
We want to see how defects influence the geometric structure of
the medium. Equation~(\ref{eq:def1}) states the relation between
the cartesian coordinates $x^i$ of a physical point in the medium
and the new cartesian coordinates ${x'}^i$ after the deformation.
This correspondence is of the form ${x'}^i={x'}^i(x^j)$. The
change in the distance described in Eq.~(\ref{eq:def2}) can be
thought of as a change in the metric tensor $g_{ij}$, which
defines the distances between the physical points. In terms of the
metric tensor we can write
\begin{equation}
dl^2=g_{ij}dx^idx^j \rightarrow d{l'}^2=g'_{ij}dx^idx^j .
\label{eq:eqdl2}
\end{equation}
{}From Eq.~(\ref{eq:def2}) we obtain
\begin{equation}
g_{ij} \rightarrowtail g'_{ij}=g_{ij}+2\varepsilon_{ij}.
\label{eq:ggprime}
\end{equation}
In this description, the coordinates of each physical point of the
medium, before and after the deformation, are referred to a
cartesian system of coordinates, which constitutes the background
of the deformation. The coordinate lines of this system coincide
with the lattice of the medium before the deformation, but things
are different  when the deformation has taken place.

This point of view is typical of a hypothetical external observer
who is able to measure distances between two physical points of
the medium and can confront the changes that occur with respect to
a fixed external background.  However, the distances can change
not only because of the presence of defects, corresponding to an
incompatible deformation. An external observer is able to notice
the change even when the deformation is compatible (such is the
case for a global dilatation). The external point of view is
misleading, because the change in the metric is not directly
linked to the presence of defects. It is more convenient to use a
more intrinsic, that is, geometric point of view.

A compatible deformation can be described using another set of
coordinates. Let us imagine that during the deformation the
coordinates are dragged with the medium. Each particle of the
medium is identified by a triplet of coordinates that does not
change during the deformation. This identification is expressed by
\begin{equation}
y^{k'}\, \hat =\,\, \delta_{k}^{k'}x^k, \label{eq:trastrasc}
\end{equation}
which means that in the new system of coordinates, each particle
has the same coordinates that it had before in the the ordered
lattice. These coordinates are called ``intrinsic''(see
Ref.~\onlinecite{bib:kron}).

An ``internal'' observer measures distances and labels  and counts
particles.  Nothing changes when the number of particles does not
change and the labelling is the same, as expressed by
Eq.~(\ref{eq:trastrasc}). The internal observer uses the
coordinates $y^{k'}$ as if they were Euclidean, although they are,
in general, curvilinear if referred to the external cartesian
background.

Things are different for an incompatible deformation. The internal
observer notices a change in the number of particles along a round
trip in the medium. There are excess holes or particles, and the
labelling is not as simple as in Eq.~(\ref{eq:trastrasc}). These
intrinsic coordinates, together with the point of view of an
internal observer, are useful for finding an incompatible
deformation, that is, the presence of defects. The new and
interesting thing is that the point of view of the internal
observer suggests that we follow a purely geometric approach for
describing the state of the medium. We shall see that the presence
of defects is seen by the internal observer in terms of a change
in the geometric properties of the medium. The internal geometry
will no longer be Euclidean.

The intrinsic view suggests that relations can be found between
the geometric properties of the medium (thought of as a
differential manifold) and the densities of defects that influence
them. These relations must be tensorial, that is, they must have a
coordinate-independent meaning.

Suppose that our internal observer can perform measurements, that
is, she has standard rods, and, moreover, is a good geometer: she
knows how to transport rods and, in general, vectors and tensors
in the medium. The result of the parallel transport of a vector
$v^h$ along an infinitesimal displacement $dx^m$ is an
infinitesimal change in the vector itself\cite{bib:ruffi}:
\begin{equation}
dv^h=-\Gamma_{ml}^{h}v^l dx^m ,\label{eq:trparcont}
\end{equation}
where $\Gamma_{ml}^{h}$ are the connection coefficients. In the
absence of defects, the medium is Euclidean and cartesian
coordinates are used and the connection coefficients are zero,
while if there are defects, the observer notices that the ``space
of the medium'' is not Euclidean, because the result of parallel
transport along a closed contour is not zero. We want to link this
idea to a geometric object, that is a tensor, so that its
properties will not depend on the specific choice of the
coordinates in the medium, but they will represent a real
geometric structure. It is clear that the connection coefficients
are somehow related to the defects, and the easiest way to see
this relation is to give a new interpretation of the Burgers
vector that we introduced before.

\begin{figure}[h]
\begin{center}
\includegraphics[width=12cm,height=6cm]{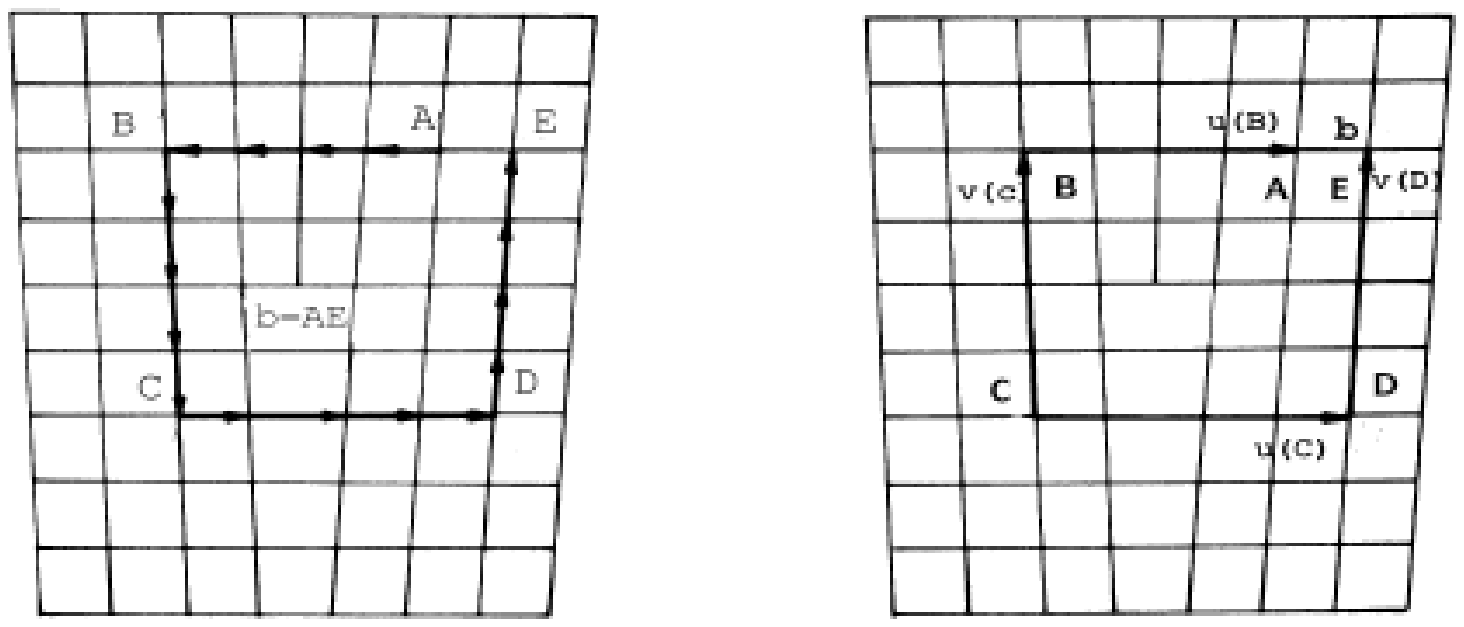}
\caption{Dislocations and geometry: (a), on the left the usual
Burgers circuit around an edge dislocation and; (b), on the right,
the differential geometric view in terms of parallel transport.}
\label{fig:kro292}
\end{center}
\end{figure}

In Fig.~\ref{fig:kro292}(a) we see the closure around a
dislocation line represented by the Burgers vector. In
Fig.~\ref{fig:kro292}(b)  we see a different procedure, which has
an intrinsic geometric meaning.\cite{bib:kron} The vectors
$\mathbf{u}(B)$ and $\mathbf{v}(D)$ are obtained, respectively, by
parallel transporting $\mathbf{u}(C)$ along $\mathbf{v}(C)$ and
$\mathbf{v}(C)$ along $\mathbf{u}(C)$. The gap we obtain is the
Burgers vector. This procedure can be described using a continuous
approach if we use the infinitesimal vectors $d_1x^h(C)$ and
$d_2x^h(C)$ (1 and 2 refer to the coordinate lines, so that
$\mathbf{u}(C)$ is parallel to direction 1 and $\mathbf{v}(C)$ is
parallel to direction 2). By parallel transporting, we obtain
\begin{subequations}
\label{eq:tp}
\begin{eqnarray}
d_1 x^h (B) & = & d_1 x^h (C) - \Gamma_{ml}^{h}d_1 x^l d_2 x^m
\label{eq:tp1} \\
d_2 x^h (D) & = & d_2 x^h (C) - \Gamma_{ml}^{h}d_2 x^l d_1 x^m,
\label{eq:tp2}
\end{eqnarray}
\end{subequations}
and the closure gap is
\begin{equation}
d_2 x^h (C)+d_1 x^h (B) - d_1 x^h (C) - d_2 x^h (D) =
\Gamma_{ml}^{h} dA^{ml} , \label{eq:difchius}
\end{equation}
where $dA^{ml}=d_1 x^m d_2 x^l - d_1 x^l d_2 x^m$ is the
antisymmetric tensor associated with the area delimited by these
vectors. Because of the antisymmetry of $dA^{ml}$, the symmetric
part of the connection coefficient is excluded in
Eq.~(\ref{eq:difchius}), and the remaining antisymmetric part,
$S_{ml}^{h}=\Gamma_{[ml]}^{h}$, is called the torsion tensor. Note
that the connection coefficients are not tensors, while their
antisymmetric part is a tensor. From the relation
\begin{equation}
db^h=S_{ml}^{h}dA^{ml}, \label{eq:dbT}
\end{equation}
we can interpret $db^h$ as the infinitesimal closure defect (or
the infinitesimal Burgers vector), and relate it to the torsion
tensor. Equation~(\ref{eq:dbalpha2}) can be written in the form
\begin{equation}
db^h=\frac{1}{2}\alpha^{nh} e_{nml}dA^{ml} \doteq
\alpha_{ml}^{h}dA^{ml}, \label{eq:dbalpha3}
\end{equation}
where $e_{nml}$ and $e^{nml}$ are defined in terms of the
completely antisymmetric tensors $\epsilon_{nml},\epsilon^{nml}$,
and the determinant of the metric $g$:
\begin{equation}
e_{nml} = \sqrt{g}\epsilon_{nml}\,\ e^{nml} =
\frac{1}{\sqrt{g}}\epsilon^{nml}. \label{eq:epsilon}
\end{equation}
If we compare Eqs.~(\ref{eq:dbT}) and (\ref{eq:dbalpha3}), it is
easy to obtain
\begin{equation}
S_{ml}^{h}=\alpha_{ml}^{h} . \label{eq:Talpha3}
\end{equation}
The meaning of Eq.~(\ref{eq:Talpha3}) is that dislocations,
through their density $\alpha_{ml}^{h}$ (defined in
Eq.~(\ref{eq:dbalpha3})), constitute the sources for torsion. This
result is very general, and does not depend on the coordinates we
used, because torsion is a tensor.

We have learned how dislocations modify the geometry of the
medium, but dislocations are not the only effect of defects.  The
curvature of the medium also is influenced by the presence of
defects. From a geometric point of view, it is known that in the
presence of curvature, a vector undergoes a nonzero change when
parallel transported along a closed path.\cite{bib:klein89} If the
path is small and encloses a small surface characterized by
$\Delta A^{nm}$, the change of the vector $v^h$ is given by
\begin{equation}
\Delta v^h = -\frac{1}{2}R_{nml}^{h}\Delta A^{nm}v^l,
\label{eq:varv^h}
\end{equation}
where $R_{nml}^{h}$ is the curvature tensor, which is
antisymmetric in the $nm$ couple of indices.

Both torsion and curvature are manifested when parallel transport
in the manifold is performed. They are global properties because
locally the space can always be assumed to be flat. Generally
speaking, we can imagine that, in the presence of defects, there
is a source term for the curvature tensor
\begin{equation}
R_{nml}^{h}=\Theta_{nml}^{h}, \label{eq:RTheta}
\end{equation}
but the meaning of the curvature in the theory of defects is more
understandable if we use the (three-dimensional) Einstein tensor
$G_{ij}$, which is a contraction of the curvature tensor
\begin{equation}
G^{ij} \equiv \frac{1}{4}e^{inm} e^{jlq}R_{nmlq} .
\label{eq:einriem}
\end{equation}
Moreover, in a linear theory of defects (see Sec.~\ref{sec:TD}),
it is easy to show that the incompatibility equation
(\ref{eq:incII}) can be written as
\begin{equation}
\chi^{ij}=G^{ij}-\partial_k \epsilon^{jkl}k_l^{i},
\label{eq:incein}
\end{equation}
where
\begin{equation}
k_l^{i}=(-\alpha_l^{i}+\frac{1}{2}\delta_l^{i}\alpha^n_{n}).
\label{eq:kappacon}
\end{equation}
If we put the source terms on the same side, we can also write
Eq.~(\ref{eq:incein}) as
\begin{equation}
G^{ij}=\sigma^{ij},
\end{equation}
where $\sigma^{ij}=\chi^{ij}+\partial_k
\epsilon^{jkl}(-\alpha_l^{i}+\frac{1}{2}\delta_l^{i}\alpha^n_{n})$.

We have introduced dislocations and disclinations, both from a
mathematical and pictorial point of view. We then described the
state of the medium by means of differential geometry and,
finally, we have related the geometric properties of the medium to
the defects, which are sources for the nontrivial structure of the
\textit{continuum} we consider.  The next step is to study the
correspondence with ECT, but before doing so, we must review this
theory.

\section{The Einstein-Cartan Theory of Gravitation} \label{sec:ECT}
In the general theory of relativity, gravitation is explained in
terms of the geometric properties of space-time. Space-time itself
is seen as a dynamical object whose structure is determined by the
energy-momentum distribution, and which determines the motion of
the bodies contained in it.

To stress the role of geometry, Wheeler called the general theory
of relativity ``geometrodynamics.''\cite{bib:MTW} Up to now,
Einstein's theory, besides having an intrinsic elegance, has
satisfied all the observational tests (see
Ref.~\onlinecite{bib:will}), so that it is the classic and
commonly accepted theory of gravitation. The effects of
gravitation are determined by mass and its motion only in the form
of energy and momentum. Of course, energy of different origin,
such as electromagnetic energy, also influences the geometry of
space-time. The constituents of macroscopic matter are elementary
particles, which obey, at least locally, quantum mechanics and
special relativity. As a consequence, elementary particles can be
classified by the irreducible unitary representations of the
Poincar\'{e} group and can be labelled by mass and spin. However,
spin has no role in general relativity, that is, spin does not
influence the geometry of space-time. This lack of influence is
not a surprise if we confine ourselves to the study of macroscopic
distributions of matter. Although mass has a monopole character
(which is additive), spin is dipolar, and its effects are
macroscopically zero, or averaged to zero (we are speaking of
unpolarized spins). But this lack of influence is not completely
satisfactory, because it is not clear why the spin cannot have a
role in determining the geometry of the space-time continuum.
Moreover, it cannot be excluded that, at the microscopical scale,
spin could play a role.

The task of finding an agreement between a theory of gravitation,
such as general relativity, and the theory of elementary
particles, to obtain a quantum theory of gravitation, is one of
the current problems of research. From a purely geometric point of
view, general relativity can be amended to include spin to
determine the properties of space-time. This is done in the
Einstein-Cartan theory, which is based on a generalization of the
geometric structure of general relativity. The Riemannian
curvature was generalized in the 1920's by
Cartan,\cite{bib:cartan22,bib:cartan86} who introduced torsion
degrees of freedom. From the late 1950's to the early 1960's, the
concept of torsion was included in the formulation of gravitation
as a gauge theory of the Poincar\'{e} group by
Kibble\cite{bib:kibble} and Sciama.\cite{bib:sciama} The theory of
gravitation with torsion is called the Einstein-Cartan theory, and
it is described extensively in Ref.~\onlinecite{bib:hehl76}, where
it is shown that the ECT is the local gauge theory of the
Poincar\'{e} group in space-time. Other details are worked out by
Hehl\cite{bib:hehl73b,bib:hehl74} and by De Sabbata and
Gasperini.\cite{bib:gaspa} A more recent review has been done by
Hammond.\cite{bib:hammond} We shall give a very introductory
primer to this theory and stress the differences with Einstein's
theory.

Mathematically, ECT differs from general relativity because
torsion and (not only) the metric shape the structure of
space-time. Physically, the presence of torsion is determined by
the intrinsic spin of matter. Although energy-momentum is coupled
to the metric tensor (and hence to curvature), spin is coupled to
torsion. We have illustrated the geometric intuitive meaning of
torsion, which is related to the closure gap of parallelograms.
Now we have to transpose everything to four dimensions, but the
idea remains the same. In the Einstein-Cartan theory of
gravitation, space-time is defined as a semi-Riemannian manifold,
endowed with a connection compatible with the metric, called a
Riemann-Cartan space ($U_4$). The general form of a connection
compatible with the metric is,\cite{footnote4}
\begin{equation}
\Gamma_{\alpha \beta}^{\gamma}=\{_{\alpha \beta}^{\gamma}
\}+K_{\alpha \beta}^{\gamma}, \label{eq:metricita12}
\end{equation}
where $\{_{\alpha \beta}^{\gamma} \}$ are the Christoffel symbols
(which are symmetric in the lower indices) and $K_{\alpha
\beta}^{\gamma}$ is the \textit{contortion} tensor and is a linear
combination of torsion tensors (see Ref.~\onlinecite{bib:hehl76}).
Because $K_{\alpha \beta}^{\gamma}$ is not symmetric in the lower
indices, we see that the ``gammas'' in Eq.~(\ref{eq:metricita12})
are not symmetric, while they are symmetric in general relativity,
where, in particular, they reduce to the Christoffel symbols.

Einstein's field equations determine the structure of space-time,
relating Einstein's tensor $G_{\alpha \beta}$, which is a
combination of the derivatives of the metric, to the
energy-momentum tensor $T_{\alpha \beta}$:
\begin{equation}
G_{\alpha \beta}=\kappa T_{\alpha \beta}, \label{eq:eife}
\end{equation}
where $\kappa=8\pi G/c^4$ is Einstein's gravitational constant.

The presence of torsion introduces new degrees of freedom in ECT,
so that we write two field equations:\cite{bib:klein89}
\begin{eqnarray}
G^{\alpha \beta}-\frac{1}{2}D^*_\gamma(P^{\alpha
\beta\gamma}-P^{\beta \gamma\alpha}+P^{\gamma \alpha \beta}) & = &
\kappa T^{\alpha \beta}
\label{eq:campoEC10} \\
P_{\alpha \beta}^{\gamma} & = & \kappa \Sigma_{\alpha
\beta}^{\gamma}. \label{eq:campoEC11}
\end{eqnarray}
where $D_\gamma$ is the covariant derivative,
$D^*_\gamma=D_\gamma+2S_\gamma=D_\gamma+2S_{\gamma
\lambda}^{\lambda}$, and $P_{\alpha \beta}^{\gamma}$ is the
Palatini tensor ($\frac{1}{2}P_{\alpha \beta}^{\gamma} \equiv
S_{\alpha \beta}^{\gamma}+\delta_{\alpha}^{\gamma}S_\beta -
\delta_{\beta}^{\gamma}S_\alpha$).

In Eqs.~(\ref{eq:campoEC10}) and (\ref{eq:campoEC11}), together
with the energy-momentum tensor $T_{\alpha \beta}$, spin is
introduced by the spin-current-density tensor $\Sigma_{\alpha
\beta}^{\gamma}$. We see that Eq.~(\ref{eq:campoEC10}) generalizes
Eq.~(\ref{eq:eife}), while Eq.~(\ref{eq:campoEC11}) is a new
equation for torsion. It is interesting to note that
Eq.~(\ref{eq:campoEC11}) is an algebraic equation: torsion does
not propagate far from its sources. It follows that the field
equations reduce to Einstein equations in vacuum. Because
practically all tests of general relativity are based on
consideration of Einstein's equations in empty space, there is no
difference in this respect between general relativity and ECT: the
latter is as viable as the former.

It is expected that differences are present inside a spin
distribution, and it is interesting to evaluate when the effects
of spin are comparable to the mass effects. To evaluate the spin
effects, we write Eq.~(\ref{eq:campoEC10}) in a different
form:\cite{bib:hehl76}
\begin{equation}
G^{\{\} \alpha\beta} = \kappa \widetilde{T}^{\alpha \beta},
\label{eq:campoeff}
\end{equation}
where $G^{\{\} \alpha\beta}$ is the symmetric (that is,
Riemannian) part of the Einstein tensor (see
Appendix~\ref{sec:appa}), and $\widetilde{T}^{\alpha\beta}$ is an
effective energy momentum tensor, which has the form (leaving
indices aside):
\begin{equation}
\widetilde{T}=T+\kappa \phi^2 , \label{eq:eq:campoeff1}
\end{equation}
where $T$ is the usual energy momentum tensor and $\phi$ is a
tensor expressing the contribution of spin. It is clear that the
effects of torsion are comparable to those of curvature when $T
\simeq \kappa \phi^2$. In particular, if the mass density is
$\rho=nm$, where $n$ is the number density and $m$ is the particle
mass, and $s=n \hbar/2$ is the spin density, we expect spin
effects to be of the same order as the mass effects when $\bar{n}
\simeq m/\kappa \hbar^2$, or, alternatively, when the matter
density is $\simeq 10^{47}$\,g\,cm$^{-3}$ for electron-like matter
and $10^{54}$\,g\,cm$^3$ for nucleon-like matter.\cite{bib:hehl76}
These are extremely high densities, which are never reached in
normal situations, even in extreme astrophysical objects. However,
while in normal conditions the effects of torsion are completely
negligible, they are expected to be important in cosmology.

Trautman\cite{bib:traut99} introduced a characteristic length to
estimate the effects of torsion, the ``Cartan'' radius. To achieve
the condition $\rho \simeq\kappa s^2$, we can imagine that a
nucleon of mass $m$ should be squeezed so that its radius
coincides with the Cartan radius $r_{\rm Cart}$:\cite{bib:traut99}
\begin{equation}
\frac{m}{r^3_{\rm Cart}} \simeq \kappa \bigl(\frac{\hbar}{r^3_{\rm
Cart}} \bigr)^2, \label{eq:rcartan}
\end{equation}
whence
\begin{equation}
r_{\rm Cart} \simeq (l^*)^{\frac{2}{3}}(r_{\rm
Compt})^{\frac{1}{3}}, \label{eq:rcart}
\end{equation}
where $l^*\simeq 1.6\times 10^{-33}$\,cm is the Planck length, and
$r_{\rm Compt}$ is the Compton length. For a nucleon we obtain
$r_{\rm Cart}\simeq 10^{-26}$\,cm, which is very small when
compared with macroscopical scales, but it is larger than the
Planck length. Hence,  torsion must be taken into account to
achieve a quantum theory of gravity.

Is it possible to detect the effects of torsion? As
Hammond\cite{bib:hammond} has carefully pointed out, the main task
is to separate potential new torsion effects from the ones
explained by known forces and fields.  Moreover, even if from a
geometric viewpoint  torsion is well defined, there are various
formulations of its generation from different sources, which make
the search for the observed effects very difficult. To detect the
effects of torsion, laboratory tests have been conceived such as
trying to exploit an alignment of spin resulting in appropriate
materials or studying the behavior of gyroscopes near torsion
sources. On the large scale, there are possible implication of
torsion on the evolution of the universe. For example, it has been
shown that a spin fluid model could prevent the big bang
singularity, even though at the same time, other models producing
torsion would enhance the singularity (see
Refs.~\onlinecite{bib:hehl76}~and~\onlinecite{bib:hammond}). More
recently, torsion has been incorporated in cosmic string theory,
and as we shall point out, this way can be directly related to our
analogy between the theory of defects and ECT. For a wide and
careful analysis of torsion, both from a theoretical and
experimental point of view, we refer to the excellent paper by
Hammond.\cite{bib:hammond}

\section{Comparison of the Theory of Defects and the Theory of
Einstein-Cartan}\label{sec:TDECT} In the previous sections we have
outlined the main features of theory of defects and ECT. Now we
will show the similarities and the differences between the two
theories. Before we continue, we want to stress the underlying
hypotheses that we assume. It is obvious that the two theories
deal with different subjects, and we expect that any
correspondence should be considered only as an analogy, and no
more than that.

The basis of this analogy are the geometric methods that are used
in both theories, because, although they are applied in a
different geometric context, namely, the three-dimensional space
for the theory of defects and the four-dimensional space-time for
ECT, they both refer to continua. Thus, the two theories have a
common formalism, and we look for analogies in the geometric
objects that characterize the two theories. For instance, torsion
plays a fundamental role in both theories.

The Einstein-Cartan theory is a geometric theory. Thus, we can use
its field equations and geometric objects in a three-dimensional
context. Consider the Einstein-Cartan field equations,
Eqs.~(\ref{eq:campoEC10}) and (\ref{eq:campoEC11}). In three
(spatial) dimensions, they can be written using a linear
approximation in the form:
\begin{eqnarray}
G_{ij}-\frac{1}{2}\partial_k(P_{ij}^{k}-P_{j\,i}^{k}-P^k_{ij}) & =
& \kappa T_{ij} \label{eq:campo10} \\ P_{ijk} \equiv
2S_{ijk}+2\delta_{ik}S_j-2\delta_{jk}S_i & = & \kappa
\Sigma_{ijk}. \label{eq:campo11}
\end{eqnarray}
In three dimensions, $T_{ij}$ and $\Sigma_{ijk}$ coincide,
respectively, with the force stress field and the moment stress
field.

After a few calculations (see Ref.~\onlinecite{bib:klein89}), it
can be shown, using the relations between the torsion and
dislocation density given in Eq.~(\ref{eq:Talpha3}), that the
field equation (\ref{eq:campo11}) becomes
\begin{equation}
P_{ijk}=\epsilon_{ijl}\alpha_{k}^{l}, \label{eq:campo12}
\end{equation}
which lets us identify the moment stress tensor with the density
of dislocations: $\epsilon_{ijl}\alpha_{k}^{l} = \kappa
\Sigma_{ijk}$. After similar calculations, the field equation
(\ref{eq:campo10}) reduces to the incompatibility equation
(\ref{eq:incein})
\begin{equation}
\chi^{ij}=G^{ij}-\partial_k \epsilon^{jkl}k_l^{i},
\label{eq:inceinbis}
\end{equation}
where $k_{li}$ has been introduced in Eq.~(\ref{eq:kappacon}).
Equation~(\ref{eq:inceinbis}) allows us to identify the total
defect density tensor with the force stress tensor, $\chi_{ij}=k
T_{ij}$.

Thus, we have seen that, formally, the Einstein-Cartan field
equations describe the defect state of a three-dimensional
continuum, at least when the defects are small so that we can use
a linear approximation. The analogy is completed by the
conservation equations, which, stated as geometric identities,
give the correct conservation laws for dislocations and
disclinations.

This analogy should not be surprising, because, as we said before,
it is based on the common geometric structure of the two theories.
In particular, it is evident that the comparison cannot be done
with general relativity, where the torsion is zero.

Now we ask if we start from this three-dimensional correspondence
in the linearized theory, is it possible to say something about
the $3+1$ space-time situation? In other words, we want to
investigate under what assumptions we can treat space-time as a
defect state of a four-dimensional continuum. The bases of our
investigation are the geometric analogies. Let us point out the
main features of these analogies.

When passing from 3 to 4 dimensions, there is an important
difference in the geometric description of the medium. In three
dimensions we can say that the effect of disclinations is to
produce curvature. We used the Einstein tensor $G_{ij}$ to write
the incompatibility equation, but we did not say explicitly that
the curvature tensor and the Einstein tensor are equivalent.  It
is well known that in three dimensions they have the same number
of independent components, which means that when the curvature
tensor is zero, the Einstein tensor also is zero and vice versa.
The presence of defects produces a nontrivial Einstein tensor,
which also means that the curvature tensor is not zero. The same
correspondence does not hold in four (or more) dimensions, because
we can have curvature even if the Einstein tensor is zero.  In
particular, far from the sources, the curvature tensor could be
nonzero. Indeed, this happens also in general relativity, because
space-time is curved even far from the sources.

If we extend the analogy to a four-dimensional context, we should
expect that the effects of defects propagate through the manifold
and are not purely local, as in three dimensions.

Kleinert,\cite{bib:klein89} adopted a linearized approach and
showed that space-time with torsion and curvature can be generated
from a flat space-time using ``singular coordinate
transformations,'' and is completely equivalent to a medium filled
with dislocations and disclinations. In other words his singular
coordinate transformations are the space-time equivalent of the
plastic deformations which lead to incompatible states (see
Sec.~\ref{sec:TD}). Hence, at least in this approximation,
space-time can be thought of as a defect state, and defects are
nothing but mass, mass current, and spin.

The next important point is to try to go beyond the linear
approximation. We did our previous comparisons assuming small
defects in order to use a linearized theory. As we have said, this
assumption is fundamental for defining a density of disclinations.
We must consider also that in real bodies there are physical
constraints on the size of defects: additional or missing matter
should not be such as to produce cracks in the structure. Hence,
from a phenomenological point of view, it is often  sufficient to
use a linear theory, as it usually the case for the elastic theory
of defects, where the linear Hooke's law is used.

However, curvature and torsion can always be introduced
geometrically by the parallel transport procedure that we outlined
in Sec.~\ref{sec:geometry}. No approximation is contained in the
equations governing curvature and torsion, so from the viewpoint
of a geometric treatment no linearization is needed.

If we keep in mind the differences that arise when we pass from
three to four dimensions, we can write four-dimensional equations
which characterize the state of the medium:
\begin{eqnarray}
G^{\alpha \beta} & = & \sigma^{\alpha
\beta} \label{eq:ec1} \\
S_{\alpha \beta}^{\gamma} & = &\alpha_{\alpha \beta}^{\gamma}.
\label{eq:ec2}
\end{eqnarray}
As we show in Appendix~\ref{sec:appa}, Eq.~(\ref{eq:ec1}) is
equivalent to a nonlinear generalization of the incompatibility
equation, and Eq.~(\ref{eq:ec2}) states the proportionality
between torsion and the dislocations tensors. In a
three-dimensional context these equations are equivalent
respectively to Eqs.~(\ref{eq:campo10}) and (\ref{eq:campo11}),
and they can be interpreted as constitutive equations. In other
words, Eqs.~(\ref{eq:ec1}) and (\ref{eq:ec2}) determine the
geometric structure of the medium once the sources are given.

Now it is easy to compare these equations to the Einstein-Cartan
field equations written in the form:
\begin{eqnarray}
G^{\alpha \beta} & = & \kappa \Psi^{\alpha
\beta} \label{eq:cartan1} \\
P_{\alpha \beta}^{\gamma} & = & \kappa \Sigma_{\alpha
\beta}^{\gamma}, \label{eq:cartan2}
\end{eqnarray}
where $\Psi^{\alpha\beta}$ is defined by
\begin{equation}
\Psi^{\alpha\beta} \equiv
T^{\alpha\beta}+\frac{1}{2}D^*_\gamma(\Sigma^{\alpha\beta\gamma}-\Sigma^{\beta\gamma\alpha}+
\Sigma^{\gamma\alpha\beta}). \label{eq:Eimod10}
\end{equation}
The correspondence between Eqs.~(\ref{eq:cartan2}) and
(\ref{eq:ec2}) is obtained using the definition of the Palatini
tensor given in Sec.~\ref{sec:ECT}.

We can then say that Einstein-Cartan space-time can be considered
as a defect state of a four-dimensional continuum, and the
equations that describe the dynamical properties of this continuum
correspond to the incompatibility equation and torsion source
equation for space-time. This correspondence is an interesting
analogy for the Einstein-Cartan theory.

The meaning of the analogy becomes clear on physical grounds when
we use three-dimensional equations in a linear theory of defects,
where we have seen that the dislocation density is analogous to
the moment stress tensor, and the total density of defects is
analogous to the force stress tensor.

\subsection{Space-Time Defects and cosmic
strings}\label{ssec:cosmicstring}

The approach that we have outlined has been investigated more
formally by Puntigam and Soleng,\cite{bib:puntigam} who exploited
the analogy between the theory of defects and ECT in the field of
cosmic strings. The latter are topological space-time defects and
are similar to the line-like dislocations and disclinations in
three-dimensional space which carry torsion and curvature.

Exploiting the above analogy and generalizing the Volterra
processes to $3+1$ dimensions, ``distorted space-times'' are built
in the paper by Puntigam and Soleng, starting from a Minkowski
space-time.  In this analogy, the Poincar\'{e} group, which is
defined by the semidirect product $P(10)=SO(1,3)\otimes_S T(4)$,
takes the place of the Euclidean group of $\mathbb{R}^3$. In this
case we have six kinds of disclination-like deformations, and four
kinds of dislocation-like deformations, which yield 10 different
Riemann-Cartan spaces filled with topological defects.

In this generalization, distorted space-times are locally
undeformed, just as in the three-dimensional case (see
Sec.~\ref{sec:TD}), but in order to detect the effect of
space-time defects, it is possible to use a procedure similar to
the one that led to the definition of the Burgers and Frank
vectors. Without going into detail, we only mention that a Burgers
vector $\mathbf{B} \in T(4)$ and a Frank matrix $\mathcal{G} \in
S(1,3)$ are defined by the parallel transport of a tetrad  in the
Riemann-Cartan space $U_4$  around the line-like defect region. In
this way,   a space-time with curvature and torsion is thought of
as a distorted medium filled with dislocations and disclinations,
and it is shown that the matter distributions, which in the
Einstein-Cartan theory act as sources, can be interpreted as
cosmic strings and cosmic dislocations. The analogy that we have
described seems promising for the study of such objects.

\section{Summary} \label{sec:conclusions}
The Einstein-Cartan theory of gravitation has been introduced
starting from an analogy with the static theory of defects, which
describes the equilibrium state of a three-dimensional continuum.
By introducing the fundamental equations of the Einstein-Cartan
theory, we recalled that it agrees with the known tests of general
relativity. Moreover, we stressed that the Einstein-Cartan theory
has a richer geometric and physical structure. In particular, a
non-symmetric connection is used, and torsion is linked to the
density of spin. In ECT, both the mass and spin determine the
geometric properties of space-time and shape its structure.

We also showed that in the classical theory of defects, a
geometric approach is possible and leads to the description of a
continuous medium by means of geometric entities that are
determined by the presence of defects, such as disclinations and
dislocations which we related to curvature and torsion. Then we
outlined a comparison between these two theories, which share a
similar underlying geometric structure, even though they apply to
very different physical phenomena.

We showed that the equations that describe the state of the medium
and its structure in the presence of defects may be interpreted as
the Einstein-Cartan field equations for a three-dimensional
continuum, at least in the linear approximation. On the other
hand, the incompatibility equation of the theory of defects, which
is usually obtained in the linear approximation, can be extended
to a more general situation, where defects are not necessarily
assumed to be small.

By pursuing this formal analogy, space-time, as described by ECT,
can be interpreted as a defect state of a four-dimensional
continuum. We suggest that this analogy, although formal, might be
useful in modern astrophysics, because cosmic strings may be
interpreted as extensions of three-dimensional defects.

It is fascinating that the theory of defects, whose origin dates
back to the beginning of the 20th century, can have such a strong
and fruitful analogy with recent developments in theoretical
physics. We believe that the analogy we have outlined can be
useful for understanding the key concepts of differential geometry
and the geometric theories of gravitation as well as helping to
stimulate interest in these fields.

\appendix
\section{Nonlinear generalization of the
incompatibility law}\label{sec:appa} It is easy to generalize the
incompatibility equation (\ref{eq:incII}) on purely geometric
grounds, extending the definition given in the linear
approximation to a general relation. We define
\begin{equation}
R_{nml}^{k} = \partial_n \Gamma_{ml}^{k} -\partial_m
\Gamma_{nl}^{k} + \Gamma_{np}^{k}\Gamma_{ml}^{p}
 - \Gamma_{mp}^{k}\Gamma_{nl}^{p},
\label{eq:Curv10}
\end{equation}
as the curvature tensor associated with the connection
$\Gamma_{ij}^{k}$. We obtain the Riemann tensor $R_{nml}^{\{\}k}$,
which  consists of the purely symmetric part of the connection,
and which is obtained from Eq.~(\ref{eq:Curv10}) by substituting
the Christoffel symbols $\{ _{ij}^{k} \}$ into the connection
coefficients $\Gamma_{ij}^{k}$. As we recalled in Sec.
\ref{sec:ECT}, the general form of the connection compatible with
the metric in the Riemann-Cartan spaces is
\begin{equation}
\Gamma_{jk}^{i} = \{_{jk}^{i} \} +K_{jk}^{i}, \label{eq:gammas}
\end{equation}
and although the Christoffel symbols are symmetric in the lower
indices, the contortion  tensor $K_{jk}^{i}$ is not, and hence the
connection is no longer symmetric.

If we start from Eq.~(\ref{eq:Curv10}), it is straightforward to
obtain the symmetric part of the Einstein tensor $G^{\{\} ij}$.
Because in the linear approximation, the incompatibility tensor is
related to the symmetric part of the Einstein
tensor,\cite{bib:klein89}
\begin{equation}
G^{\{\} ij} \equiv \epsilon^{ikl}\epsilon^{jmn}\partial_k
\partial_m \varepsilon_{ln},
\label{eq:inc01}
\end{equation}
the natural extension of the incompatibility equation
(\ref{eq:incII}) can be performed using Eq.~(\ref{eq:inc01}).

The source equation for the Einstein tensor,
\begin{equation}
G^{ij}=\sigma^{ij}, \label{eq:einsteinsigma}
\end{equation}
is written in the form,
\begin{equation}
G^{ij} \equiv G^{\{\} ij}-\eta^{ij}-\psi^{ij} = \sigma^{ij},
\label{eq:Gchi10}
\end{equation}
where
\begin{subequations}
\label{eq:Curv17}
\begin{eqnarray}
E_{jkl} & = & \frac{1}{2}(\partial_j \varepsilon_{kl}
+\partial_k \varepsilon_{jl}- \partial_l \varepsilon_{jk}) \\
\eta^{ij} & \equiv & -\frac{1}{2} e^{inm} e^{jlq}\partial_n
K_{mlq} \label{eq:Curv16} \\
\psi^{ij} & \equiv & \frac{1}{2} e^{inm} e^{jlq}
g^{kp}(2E_{nqk}K_{mlp}+2E_{mlp}K_{nqk}+K_{nqk}K_{mlp})
\end{eqnarray}
\end{subequations}
(see Ref.~\onlinecite{bib:gai}). Moreover, $G^{\{\} ij}$ can be
written as
\begin{equation}
G^{\{\} ij} = e^{inm} e^{jlq}\partial_n
\partial_l \varepsilon_{mq} -2 e^{inm} e^{jlq}g^{kp}E_{nqk}E_{mlp},
\label{eq:Curv18}
\end{equation}
and if we write $\Xi^{ij}=2 e^{inm} e^{jlq}g^{kp}E_{nqk}E_{mlp}$,
Eq.~(\ref{eq:Gchi10}) becomes
\begin{equation}
e^{inm} e^{jlq}\partial_n \partial_l \varepsilon_{mq} -\Xi^{ij}
-\eta^{ij}-\psi^{ij} = \sigma^{ij}, \label{eq:Incnonlin}
\end{equation}
which is the desired extension of the incompatibility equation. We
obtain the usual relation if we notice that $\Xi^{ij}$,
$\eta^{ij}$, and $\psi^{ij}$ are nonlinear contributions.

Although we obtained Eq.~(\ref{eq:Incnonlin}) for three
dimensions, our approach is based entirely on geometric
foundations and can be easily transposed into curved space-time,
which, in this sense, can be interpreted as an incompatibility
state.

\section{Operators on symmetric tensors}\label{sec:appb}

We give a different and perhaps more intuitive, interpretation of
some of the concepts we have introduced.
Kr\"{o}ner\cite{bib:kron58} has developed a continuum theory of
dislocations and disclinations using an approach that is very
similar to electromagnetic theory. Equations that give internal
stresses in terms of source functions (that is, defect densities)
are formulated and are analogous to Poisson equations. To give an
idea of this formalism, we show that the equations and the tensors
that we introduced in the main text can be defined in a different
way using the technique of differential operators on symmetric
tensors (see Refs.~\onlinecite{bib:kron} and also
\onlinecite{bib:dewit}).

We introduced the strain tensor in Eq.~(\ref{eq:strain}). A
different definition may be given using the deformation operator
\deff, which is a type of symmetric gradient:
\begin{equation}
\mathbf{\varepsilon}=\deff\, \mathbf{u} \Leftrightarrow
\varepsilon_{ij}= \frac{1}{2}(\partial_i u_j +\partial_j u_i).
\label{eq:def}
\end{equation}
In the same way, an incompatibility operator \inc\ may be
introduced, by which the incompatibility equation reads
\begin{equation}
\inc\, \mathbf{\varepsilon}=\mathbf{\chi} \Leftrightarrow
\epsilon^{ikm}\epsilon^{jln}\partial_k \partial_l
\varepsilon_{mn}=\chi^{ij}. \label{eq:inc}
\end{equation}
Finally, it is obvious that we may introduce a divergence operator
\divv, whose meaning is well known:
\begin{equation}
\divv\, \mathbf{\varepsilon} \Leftrightarrow
\varepsilon^{ij}_{,j}. \label{eq:div}
\end{equation}
The point is that the tensor operators \deff, \inc, and \divv\ are
analogous to the vector operators \grad, \curl, and \divv. It is
easy to verify that
\begin{equation}
\divv\,\inc \equiv 0 \quad \mbox{and} \quad \inc \ \deff \equiv 0,
\label{eq:did}
\end{equation}
which are the analogs of $\divv\, \curl \equiv 0$ and $\curl\,
\grad \equiv 0$.

As it is well known, a vector $\mathbf{v}$ that vanishes at
infinity can be decomposed into a gradient of a scalar function
$\varphi$ and the curl of a vector function $\mathbf{A}$:
\begin{equation}
\mathbf{v}=\grad\, \varphi+ \curl\, \mathbf{A}. \label{eq:vector}
\end{equation}
In the same way Kr\"{o}ner showed that a symmetric second rank
tensor $\mathbf{T}$ that vanishes at infinity can be decomposed
into two terms, that is, a deformation of a vector field
$\mathbf{u}$ and the incompatibility of a symmetric second rank
tensor $\mathbf{S}$:
\begin{equation}
\mathbf{T}=\deff\,\mathbf{u}+ \inc\, \mathbf{S}. \label{eq:tensor}
\end{equation}
Thus, if a symmetric tensor field has zero divergence, it can be
written as a pure incompatibility field, while, if it has zero
incompatibility, it can be written as a pure deformation field.

By starting from this analogy, Kr\"{o}ner was able to write the
equations relating the internal stress and strain to the sources
of incompatibility (the densities of defects). This procedure is
analogous to the one that is used in electromagnetism, where the
potential and fields are computed starting from the Poisson or
Laplace equations. The electromagnetic analogy is very useful and
provides a clear understanding of the problem of the link between
defects and internal strain. Moreover the analogy recalls closely
ECT, where the geometry of space-time is determined by the
energy-momentum tensor and spin current tensor. Indeed, in
Kr\"{o}ner's approach, the strain of the medium, that is, its
geometry, is determined by the distribution of defects.

\newpage

\begin{thebibliography}{99}
\bibitem{footnote1} In these sections we present the fundamental
ideas of the theory of defects, with the aim of drawing a formal
analogy with ECT. We shall follow the work of
Kr\"{o}ner,\cite{bib:kron} De Wit,\cite{bib:dewit}
Nabarro,\cite{nabarro67} Gairola,\cite{bib:gai} and
Kleinert.\cite{bib:klein89} References to earlier papers can be
found in these papers.
\bibitem{footnote2} Latin indices run from 1 to 3, and there is
no need to distinguish between covariant and contravariant ones,
because the background metric is Euclidean. However, we shall use
covariant and contravariant indices in summations; in order to use
the Einstein convention; $\varepsilon_{ij}dx^i dx^j$ means
$\sum_{i=1}^3 \sum_{j=1}^3 \varepsilon_{ij}dx^i dx^j$.
\bibitem{footnote3} The term defect refers both to a true
absence of particles and to an excess of them, with respect to the
ordered structure of the medium, which, according to common
terminology, we shall call an undeformed medium.
\bibitem{footnote4} Greek indices run from 0 to 3 and are raised
and lowered by the metric tensor $g_{\alpha \beta}$; the signature
of space-time is $(1,-1,-1,-1)$, and again, the summation
convention is used.
\end{thebibliography}
\end{document}